 \documentclass[authoryear,preprint,review,12pt]{elsarticle}

\usepackage[T1]{fontenc}

\usepackage{amssymb}
\usepackage{amsmath}

\usepackage{siunitx}
\usepackage{hyperref}

\usepackage{lineno}

\journal{Neural Networks}

\begin{document}

\begin{frontmatter}



\title{Evidence of Scaling Regimes in the Hopfield Dynamics of Whole Brain Model} 


\author[a,b]{Giorgio Gosti} 
\ead{giorgio.gosti@cnr.it}

\affiliation[a]{organization={Center for Life Nano- and Neuro-Science, Istituto Italiano di Tecnologia},
            addressline={Viale Regina Elena 201}, 
            city={Rome},
            postcode={I-00161}, 
            country={Italy}}
            
\affiliation[b]{organization={Istituto di Scienze del Patrimonio Culturale, Consiglio Nazionale delle Ricerche},
            addressline={Strada della Neve s.n.c., Via Salaria km 29.300}, 
            city={Montelibretti (RM)},
            postcode={I-00010}, 
            country={Italy}}
            
\author[a,c]{Sauro Succi}

\affiliation[c]{organization={Istituto per le Applicazioni del Calcolo del Consiglio Nazionale delle Ricerche},
            addressline={via dei Taurini 19}, 
            city={Roma},
            postcode={I-00185}, 
            country={Italy}}

\author[a,d]{Giancarlo Ruocco\corref{cor1}}
\ead{giancarlo.ruocco@roma1.infn.it}

\affiliation[d]{organization={Dipartimento di Fisica, Universit\'a di Roma ``La Sapienza''},
            addressline={P.le Aldo Moro 5}, 
            city={Roma},
            postcode={I-00185}, 
            country={Italy}}

\cortext[cor1]{Corresponding author}

\begin{abstract}
It is shown that a Hopfield recurrent neural network exhibits a scaling regime, whose specific
exponents depend on the number of parcels used and the decay length of the coupling strength.
This scaling regime recovers the picture introduced by Deco \emph{et al.}, 
according 
to which the process of information transfer within the human brain shows spatially 
correlated patterns qualitatively similar to those displayed by turbulent flows, although with
a more singular exponent, $1/2$ instead of $2/3$.
Both models employ a coupling strength which decays exponentially with 
the Euclidean distance between the nodes, informed by experimentally derived 
brain topology. 
Nevertheless, their mathematical nature is 
very different,  Hopf oscillators versus a Hopfield neural network, respectively.  
Hence, their convergence for the same data parameters, suggests  
an intriguing robustness of the scaling picture.
Furthermore, the present analysis shows that the Hopfield model brain remains functional 
by removing links above about five decay lengths, corresponding to about one sixth of the size of the global brain. 
This suggests that, in terms of connectivity decay length, the Hopfield brain functions in a sort of intermediate 
``turbulent liquid''-like state, whose essential connections are the intermediate ones between the connectivity
decay length and the global brain size.  
The evident sensitivity of the scaling exponent to the value of the 
decay length, as well as to the number of brain parcels employed, 
leads us to take with great caution 
any quantitative assessment regarding the specific nature of the scaling regime.
\end{abstract}


\begin{highlights}
\item The mesoscale Hopfield model of the global brain collective dynamics presents a scaling regime. 
\item It recovers the scaling picture recently introduced by Deco et al. 2020 based on a Hopf oscillator model.
 The near-quantitative convergence of the two models points to an intriguing robustness of the scaling picture. 
\item  A scaling regime implies that the 
brain functions in a sort of intermediate  ``turbulent liquid''-like state.
\item It also shows that the scaling exponent associated with such collective patterns is highly sensitive to the decaying length of the brain connectivity, as well as to the number of brain parcels.
\item The global brain Hopfield model is shown to remain functional upon removing connections
above five decay length scales.
\end{highlights}


\end{frontmatter}



\section{Introduction}
\label{sec:intro}

In  a recent paper, \cite{Deco2020} argued that the information transfer within the human brain 
may proceed in close analogy with the mechanisms which govern mass and energy transport in turbulent fluids. The analogy refers to the emergence of correlated spacetime patterns which would facilitate 
and possibly optimize information transfer across the brain, in the ``same'' way as fluid 
turbulence enhances energy transport across fluids.  
This is interesting in many respects, not least the prospects for future neurocomputers
\citep{Schuman2022}.

Based on an extended dynamical system of oscillators (Hopf model),  supplemented by large-scale 
neuro-imaging empirical data on the connections between different brain areas, 
\cite{Deco2020} numerically showed that the statistical correlations of human brain 
signals may exhibit a scaling regime similar to the one observed in turbulent fluids. 

At a closer scrutiny,  though, the analogy does not appear to be 
quantitative,  meaning that  the spatial scaling exponents of the second order 
structure function $S_2(d)$=$\langle(O(d)-O(0))^2 \rangle$$\sim d^{\alpha}$,  $O$ being a suitable 
order parameter, are not the same,  about $\alpha$$ \sim $1/2 for the Hopf model 
of the human brain against $\alpha$=2/3 for fluid turbulence \citep{Benzi1993}. 
Hence ``brain turbulence'' appears to be more singular than fluid turbulence.
 
Yet, a transient scaling regime in space is found in either cases,  
which is intriguing {\it per se}, regardless of the specific value of $\alpha$. 

In this paper, we examine the robustness of the aforementioned analogy by conducting a similar analysis based on the same neuroimaging data for brain connectivity, but using a very different model: a Hopfield neural network. 
Additionally, we measure the scaling exponents as a function of the connectivity decay 
length ($\delta$), discovering a much richer scenario, including a dependence on 
the number of parcels (network nodes) ($N$), to be detailed later in the paper. 

The recurrent Hopfield neural network  model is a well-established model for small-scale
neuronal networks \citep{Amari1972, Little1974, Hopfield1982} 
and a particularly minimal discrete-time recurrent neural network model \citep{Grossberg1967, Rumelhart1986}. 
It is a neural network model
based on binary McCulloch-Pitts neurons \citep{Amit1985,Brunel2016,Hillar2018,Hillar2021} in which every processing unit ($i$)
is connected to all the other ones ($j$) through a set of weights ($J_{ij}$). Most of the literature
considers fully connected and symmetric Hopfield networks since this simplifies
the mathematical analysis of the model, although recently the investigation of
diluted networks \citep{Stauffer2003, Brunel2016, Kim2017, Folli2018, Leonetti2020} and asymmetric networks \citep{Gopalsamy1994, Xu1996,Chen2001,Franca2000, Zheng2010, Szedlak2014, Folli2018, Leonetti2020} was introduced.

Usually, a recurrent Hopfield neural network is trained to store patterns or memories as steady-state attractors with
the Hebbian prescription. In this case, the coupling matrix $J_{ij}$ takes the form of the sum of dyadics, and 
therefore it is symmetric and fully connected. 
The maximum storage capacity of such a network is found to scale linearly with the number of nodes $N$. 
The capacity can be increased by the process of learning (Hebbian learning) whereby the simultaneous activation 
of two nodes reinforces their connection. This is inspired by biological neural networks where the simultaneous activation of neurons leads to increments in synaptic strength \citep{Hebb1949}, and it allows the network to store a 
larger number of patterns as steady-states \citep{Amit1985}. Yet, this quantity ($C$) remains linear with $N$: $C \sim N$, being the proportionality factor of the order of unity. 
Lately, researchers have also investigated how unlearning methods can improve the stability of the stored patterns \cite{Fachechi2019,Benedetti2022,Benedetti2023,Agliari2019,Agliari2024}. 
It is worth emphasizing that a vast
literature, tracing back to the seminal papers by \cite{Gardner1986} and \cite{Hopfield1983},
treat the stationary points of the Hopfield dynamics as ``memories'', but in the context of 
whole brain dynamics, and in our work, we will not consider this interpretation.

Interestingly, it has been recently demonstrated both numerically \citep{Folli2018, Gosti2019} and theoretically \citep{Hwang2019,Hwang2020} 
that {\it random}, non-Hebbian, coupling matrices have a number of steady-states 
much larger than the Hebbian ones. Indeed the number of steady-states 
$C$ of random matrices scales exponentially with $N$: $C \sim   e^{\Sigma N}$ being $\Sigma$ the complexity. The latter quantity is found to be maximized for asymmetric and diluted matrices, with values 
that match the connectivity found in the hippocampus areas.

The approach outlined by \cite{Deco2020} diverges from both the Hebbian and random connectivity models. It relies on a coupling matrix, which is derived from experimental observations of the connections between human brain regions. 
This approach assumes that the coupling strengths are all positive, deterministic, and specifically proportional to a given function of the distances between nodes. 
$J_{ij} = e^{-d_{ij}/\delta}$, $d_{ij}$ being the Euclidean distance between nodes $i$ and $j$ in the network, and $\delta$ a decay length, which, in \cite{Deco2020} are kept fix to match the empirical HCP dMRI tractography of the human brain: $\delta = 5.55 $ \si{mm}.

At variance with the latter approach, where the dynamics is based on the Hopf oscillator 
model, here we use the same coupling matrix of \cite{Deco2020}, but employ the Hopfield dynamics.

The use of Hopfield networks to investigate the collective dynamics of the 
brain activity is not new \citep{Trappenberg}. Indeed, \cite{Deco2012}, and \cite{Golos2015} used
Hopfield models (or Ising-Spin models with Glauber dynamics) 
as a neural mass model to study the complex landscape of brain activation dynamics. 
Neural mass models schemes do not interpret the nodes of the Hopfield network as  
single spiking neurons but rather as aggregates of neural mass (parcel), namely
mesoscale brain regions switching between active and rest states. 
In the active states, many neurons simultaneously fire while in the rest states most neurons are silent. 
These models do not aim at establishing a rigorous connection between the microscopic description
of the brain and a corresponding mesoscopic one, but posit instead a physically-informed heuristic
picture of the collective brain dynamics.

Broadly speaking, the Hopf model is based on the idea of reproducing the empirically observed 
information transfer between areas that takes place through (partial) synchronization of collective oscillations. 
In this framework, the natural collective variables describing the global brain dynamics
are mesoscale oscillators obeying the Hopf dynamics. 
The Hopfield model used in this work draws inspiration from a very different idea, namely to 
carry over the discontinuous divide between active and rest states from the microscale 
level of single neurons to the mesoscale level of neural parcels, as discussed above. 
Formally, this amounts to assuming that coarse-graining of the Hopfield model would
leave the model invariant but with a different set of weights, in our case an exponential decay. 
Such model proves useful to show that the dynamical repertoire of brain activity composed of sudden avalanches 
and bursts, can be described in terms of critical systems perturbed by noise \citep{Deco2012, Golos2015}.

Due to the significant difference between the fundamental assumptions behind these two models, 
inspecting their commonalities, as well as their points of departure,
starting from the same empirical connectivity information, is of decided interest to
improve the mesoscale description of the global brain dynamics,

Our main conclusions are summarized as follows: 

{\it i)} We confirm the existence of a 
{transient} scaling regime of the neural activity in space
for each value of the connectivity decay length  $\delta$ in the explored range, 0$\le$$\delta$$\le$10 mm; 

{\it ii)} At the $\delta$ value investigated in \cite{Deco2020} ($\sim 5.5$ mm) and for $N=1000$, 
we found $\alpha \sim 2/5$, close to the one  reported by \cite{Deco2020}\footnote{In Ref. \cite{Deco2020} the used parameter is the inverse of the decay scale, $\lambda=1/\delta$.}; 

{\it iii)} For $N=1000$ the scaling exponent $\alpha$ shows a sigmoid-like 
trend with the decay length $\delta$.
Importantly, for $N=1000$, the functional relation $\alpha(\delta)$ shows a steep 
increase in the region around  $\delta \sim 5.55$ mm, the value indicated in \cite{Deco2020}
as the physiologically relevant one from the analysis of empirical HCP dMRI tractography
with a $N=1000$ parcellation.
This means that the specific value of the decay length has a significant
impact on the scaling exponent; for instance, a slight increase to 
$\delta$=5.88 mm  would yield a scaling exponent slightly above $2/3$,  
the value associated with homogenoeus incompressible turbulence.
Conversely, a slight decrease to $\delta$=5 mm, would yield $\alpha$$\sim$0.1, a much less organized
regime. 

{\it iv)} the sigmoid dependence of $\alpha$ on $\delta$ changes with $N$, 
becoming steeper and narrower at increasing $N$. 

{\it v)} By systematically pruning the couplings below a given running threshold, $J_{ij}<J_{th}$ 
(which corresponds to setting to zero the connections above a certain distance),  
it is found that the scaling exponent 
remains largely unaffected until we prune up to  95$\%$ - 98$\%$ of the couplings.  
This suggests that  nodes at a distance $d  >  5 \delta \sim 25$ mm, do not
significantly partake to the physical mechanisms at the roots of the scaling regime.   


\section{\label{sec:HRNN} Discrete-time Hopfield Recurrent Neural Networks}

Following \citep{Amari1972, Little1974,Hopfield1982,Amit1985,Folli2017,Gosti2019,Leonetti2020,Gosti2024}, we consider a 
network of $N$-binary nodes that can be either firing or silent.
The state of each node is a binary variable $s_i$, with $i =1, 2,\ldots ,N$,
such that $s_i=-1$ or $s_i=1$, so that the network state is represented by 
a binary string $\mathbf{s}=(s_0,s_1,\ldots,s_N)$.
Neurons interact through the connectivity matrix $\mathbf{J}$, with matrix elements $J_{ij}$
following the same exponential dependence $J_{ij} = e^{-d_{ij}/\delta}$ used in
\citep{Deco2020}. 
We assume a discrete time $t$, and, at each time step $t$, the evolution of 
the neuron state $s_i(t)$ is given by the following non-linear dynamic equation,
\begin{equation}\label{eq:Hop}
s_i(t+1) = sign \left\lbrack\sum_{j=1}^N J_{ij}s_{j}(t) \right\rbrack,
\end{equation}
where the activation function $sign(x)$ is  defined to be 1 
for $x$$\ge$0, and -1 otherwise. 
Consequently, whenever the summation of the inputs on node $i$ is above zero, 
the node is active (``fires''), otherwise, it remains silent.  

We determined the couplings matrix $J_{ij}$ using Schaefer's cerebral 
cortical parcellation atlas as in \citep{Deco2020}. 
This parcellation was obtained from the analysis of a large dataset of neural activity dynamics
with a sample size of 1489 \citep{Schaefer2018}. 
This dataset is publicly available. We 
used Schaefer's parcellations with $N$ ranging from $200$ to $1000$. The Schaefer's parcel centroids are mapped onto 
the MNI152 volumetric space\footnote{Github repository dataset url \cite{Schaefer2018}: \url{https://github.com/ThomasYeoLab/CBIG/tree/master/stable\_projects/brain\_parcellation/Schaefer2018\_LocalGlobal/Parcellations/MNI/Centroid\_coordinates}}.
The code for reproducing the simulations can be found in the git hub repository 
\url{https://github.com/ggosti/HopBrain}.

\section{The Structure Factor $S_2(d)$}

We used \cite{Deco2020} adaptation of Kolmogorov’s concept of structure functions.
In turbulence, the Kolmogorov's structure factor applies to the longitudinal velocity 
but in our case it represents the activity of the parcel 
as measured through the BOLD signal in a fMRI experiment.

To compute the structure factor we use the auxiliary covariance function $B(d)$ defined as:
\begin{equation}
    B(d)= \sum_{ij} \frac{1}{{\cal{N}}(d)} \sum_{d_{ij} \in d } s_i s_j 
\end{equation}
where the function $ {\cal{N}}(d)$ represents the number of $i,j$ pairs at distance $d$. 
The notation $d_{ij} \in d$ indicates that the distance between the nodes must 
be ``close enough'' to $d$. To this aim, it should be kept in mind that in 
Schaefer's cerebral cortical parcellation atlas \citep{Schaefer2018}, $d$ was recorded with 2 mm precision, thus 
there are multiple $i,j$ pairs with the same $d$. 
Therefore we can bin the existing nodes' distance and map them to the natural
numbers, so as to use of the Kronecker delta $\Delta(ij \vert d)$, which 
takes value $1$ if $d_{ij}$ is in the bin identified by $d$, and is $0$ otherwise. 

As a result,  we have:
\begin{equation}
    B(d)= \frac{1}{{\cal{N}}(d)} \sum_{ij} s_i s_j \Delta(ij \vert d)
\end{equation}
and
\begin{equation}
     {\cal{N}}(d)=\sum_{ij} \Delta(ij \vert d)
\end{equation}
 The function $B(d)$ relates to the structure factor as follows \citep{Deco2020}:
\begin{eqnarray}
    S_2(d) 
    &=& 2\left[ B(0)-B(d) \right]
\end{eqnarray}
The Hopfield dynamics is deterministic, and since the phase space of the nodes' states 
is finite, the asymptotic trajectories must necessarily be periodic. 
Therefore, as also discussed in \citep{Folli2017}, Hopfield Recurrent Neural Networks, given 
an initial state, consistently end up into a 
specific steady state (``attractor''). 
The steady states are deterministic attractors composed of either limit cycles formed by 
repeating sequence of states, or a single stationary state.  
Given that in our case the couplings $J_{ij}$ are symmetric,
we expected theoretically, and experimentally confirmed that the steady states are always  
composed of a single fixed point.
In the absence of a further external input (``external stimuli''), and given sufficient time, the network gets stuck in an attractor and the whole activity is constant in time,  hence no time summation was taken in the previous equations. 
This is of course a very crude representation of the brain dynamics, which receives 
continuously ``internal'' and ``external'' inputs, that perturb the system
to constantly explore a landscape shaped by the collection of steady states.
 Given this context, \cite{Deco2012}, and \cite{Golos2015} show how the whole set of attractors allows us to reproduce the dynamical repertoire and the functional correlations of the brain activation dynamics.

\section*{Scaling exponents vs. the decay length}

\begin{figure}[h!] 
\centering
\includegraphics[width=0.6\columnwidth]{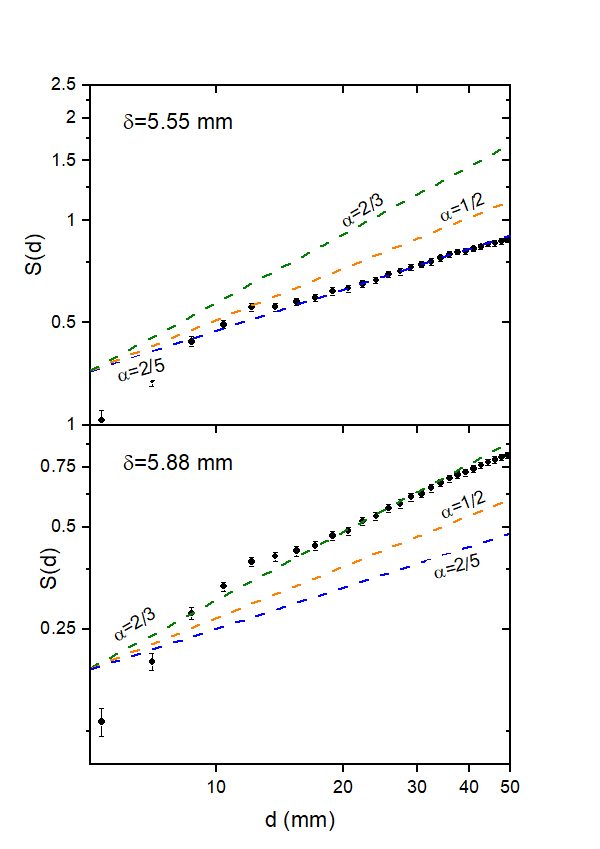}
    \caption{
    \label{fig:Sr} 
    Structure factor $S_2(d)$ for $N=1000$ nodes and $N_r=1000$ realizations 
    respectively at $\delta = 5.55$ mm (A),
    and $\delta = 5.88$ mm (B).
    The values $S_2(d)$ were binned over equal sized intervals. 
    The dashed lines indicate different $\alpha$ values: 
    $\alpha = 2/3$ which corresponds to turbulence,  $\alpha = 1/2$, the 
    value obtained in \citep{Deco2020}, and $\alpha = 2/5$. 
    We estimate the scaling exponent $\alpha$ with a linear regression of $\log(S(d)) \sim \alpha \log(d)  + \beta$  in the interval range $[2.7 - 33.1 ]$~mm, as used in \citep{Deco2020}, and we obtain 
    $\alpha \approx 2/5$ for  $\delta = 5.55$ mm, and $\alpha \approx 2/3$ for  $\delta = 5.88$ mm.
    }
\end{figure}


Given different values of the decay length $\delta$, we have run the Hopfied model for a system composed 
by $N=1000$ parcels (with pairwise distances consistent with the Schaefer's cerebral cortical parcellation atlas \citep{Schaefer2018}). The structure functions have then been calculated 
by following the system until the fixed points were reached,
and by averaging over $N_r$=1000 randomly chosen initial states. 
As examples, figure \ref{fig:Sr} 
shows 
the structure factor $S_2(d)$ for respectively $\delta$=5.55 mm (Fig. \ref{fig:Sr}A) 
and $\delta$=5.88 mm (Fig. \ref{fig:Sr}B).
The black points with the error bars represent the average $S(d)$
computed over 1000 different realizations.
The dashed green line corresponds to $\alpha$=2/3,
the orange line corresponds to $\alpha$=1/2,
and the blue line corresponds to $\alpha$=2/5.
The $\alpha$=1/2 value (orange line) is the value estimated in \citep{Deco2020}.
The $\alpha$=2/3 is the exponent associated with a turbulence.
Due to the significant level of statistical fluctuations, we have  
estimated the average exponent in two different ways,  first as the 
correct ensemble average $\langle S_2(d)\rangle$$\sim d^{\alpha}$ and also
as an arithmetic average of the single realization exponents, 
$\overline \alpha$=$N_r^{-1} \sum_{r=1}^{N_r} \alpha_r$.
Even though the latter is not correct, the fact that both yield very similar values
attests to the robustness of the average exponent,  notwithstanding sizeable fluctuations.
For consistency with \citep{Deco2020}, we estimate the scaling exponent $\alpha$ fitting the points in figure 1 in the interval 2.7 - 33.1 mm 
and we obtained $\alpha \approx$2/5 for  $\delta$=5.55 mm, and $\alpha \approx$2/3 for $\delta$=5.88 mm.

\begin{figure}[h!] 
\begin{center}
        \includegraphics[width=.75\columnwidth,trim={1.5cm 19cm 2cm .7cm},clip]{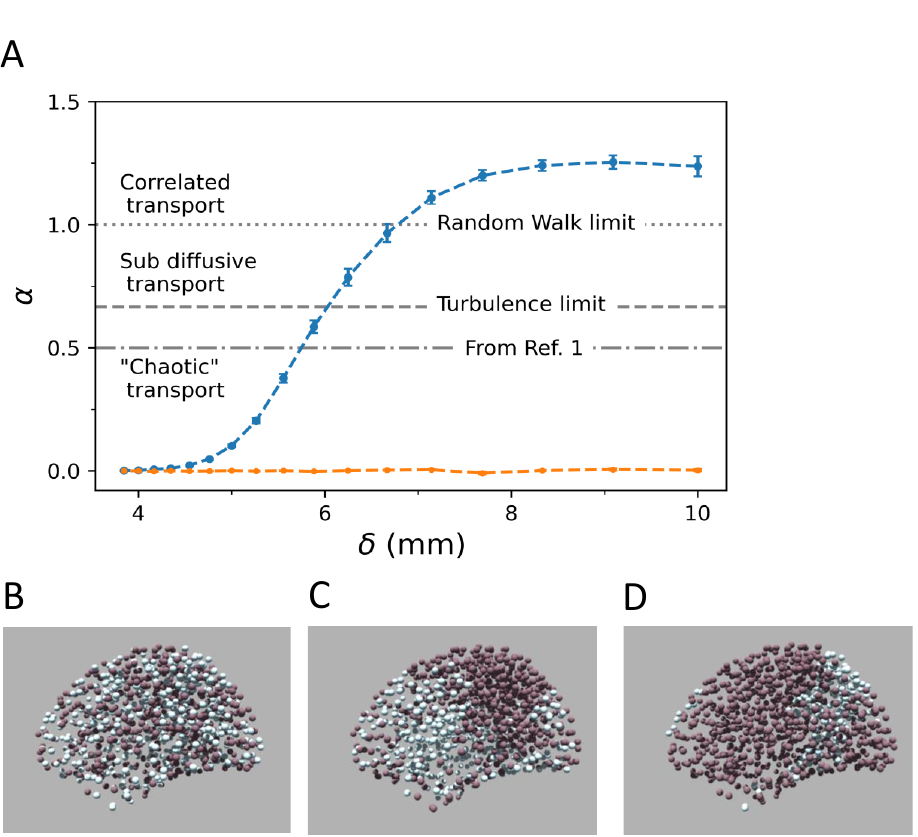}
\end{center}

\caption{\label{fig:slopeDelta}
A) The blue line shows the average slope of $\alpha = \log S(d)/ \log d$  
for different $\delta$s, given the connectivity matrix $J_{ij}=\exp{(- d_{i,j}/\delta) }$. 
The horizontal dotted lines correspond to  $\alpha_{RW}=1$ and $\alpha_T=2/3$, corresponding
to random walk and homogeneous incompressible turbulence, respectively, while the dot-dashed
horizontal line corresponds to \cite{Deco2020} 
``brain'' value $\alpha_D=1/2$.
The orange line shows the function $\alpha(\delta)$  with 
a connectivity matrix obtained by randomly shuffling all the weights $J_{ij}$, which results
into a basically uncorrelated signal, indicating the importance of the metric structure of the
weights.
The plot stops at $\delta=10$ mm,  a region where the signal 
is highly correlated ($\alpha \sim 1.2$), but still below the smooth regime 
marked by $\alpha=2$.  
Conversely, a brain below $4$ mm, would exhibit ``chaotic'' behaviour, 
with scaling exponents close to zero, \emph{i.e.} no scaling regime.
The bottom panels show the neuron activity mapped on the Schaefer’s cerebral 
cortical parcellation atlas \citep{Schaefer2018} coordinates for 
(B) $\delta = 4$ (``chaotic''), (C) $\delta=5.99$, and (D) $\delta=6.66$ (random walk),
respectively.
}
\end{figure}

We repeated the previously depicted procedure for a sequence of values of the decay length in the range $[3-10]$  mm  and  measured the associated scaling exponents. The values of $\alpha(d)$ for different realizations are shown in Fig. \ref{fig:slopeDelta} 
as blue points.  
The code to reproduce the simulation and the 3D model files (gltf) 
are available
at the repository \url{https://github.com/ggosti/HopBrain}. This code also shows how 
to generate 3D maps for the  
Schaefer parcellation's nodes activation states inspection,
and how to export gltf files
that can be viewed in immersive environments using open source webXR apps such as Aton which are part of the H2IOSC
open cloud \citep{fanini2021aton,fi16050147} (Fig. \ref{fig:slopeDelta}B-D).
Figure \ref{fig:slopeDelta}B-C can be accessed with a web-browser
respectively with the following urls:
 \url{https://aton.ispc.cnr.it/s/ggosti/20240612-6zkx4gyvi} for $\delta = 4$ (``chaotic''), \url{https://aton.ispc.cnr.it/s/ggosti/20240612-ebn1rod5p} for $\delta=5.99$, and 
 \url{https://aton.ispc.cnr.it/s/ggosti/20240612-a8e3initf} for $\delta=6.66$ (random walk).

 Several comments are in order. First, we observe a clear increasing trend of $\alpha$ with $\delta$, meaning
that small decay lengths, \emph{i.e.} localised connectivity, promotes irregular patterns.  
For instance, below $\delta \sim 4$ mm, the scaling exponent is basically 
zero, corresponding to complete randomness.
This was checked independently, by running simulations in which the 
couplings $J_{ij}$ were randomized though keeping the same distribution of the elements of the matrix $J$ (flat orange line at the bottom). We were able to simultaneously randomize the couplings $J_{ij}$, and keep the same value distribution
by randomly shuffling the $J_{ij}$ values. 
At the opposite end, with $\delta$=10 mm, we measure $\alpha$$\sim$1.2, 
which is still well below the smoothness threshold $\alpha_S$=2, but above the 
random walk value $\alpha_R$=1, and much above
the Kolmogorov turbulence value $\alpha_T$=2/3, which is in turn larger than  
$\alpha_D$$\sim$1/2, the value obtained by 
\cite{Deco2020} at $\delta$$\sim$5.55 mm. 
Importantly, in the intermediate regime, around the physiological value 
$\delta$=5.55 mm, the present Hopfield model delivers  $\alpha_H$$\sim$2/5,  different 
but still close to Deco \emph{et al.} value.
A visual extrapolation of Fig. \ref{fig:slopeDelta}A seems to indicate that the scaling regime would 
remain non-smooth ($\alpha < 2$) even in the limit of an 
"infinite-brain" ($\delta \to \infty$) the global brain ($d \sim 25 \delta$) being
pretty close to the infinite-brain limit.
Note however that 
for $\delta > 10$ mm we start getting cases in which $B(d) \sim B(0)$,  so that the 
numerical calculation of the exponent diverges and becomes less and less accurate 
(a fully correlated signal with $B(d)=B(0)$ yields formally $\alpha \to \infty$.)  

\begin{figure}[t] 
\begin{center}
\includegraphics[width=1.\columnwidth]{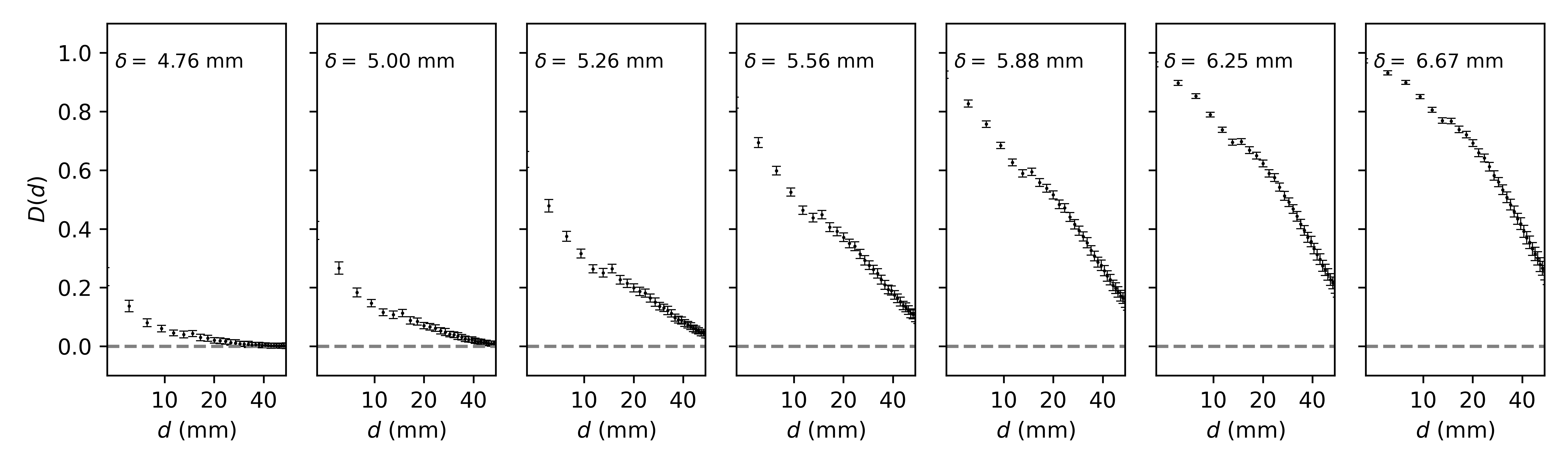}
\end{center}

\caption{\label{fig:Corr} 
Pair-correlation functions $B(d)$ for different $\delta$s.
At small $\delta$ the signal is largely uncorrelated, while at increasing 
$\delta$, correlations start to emerge,  although affected
by significant amount of statistical fluctuations.
}
\end{figure}

Figure \ref{fig:Corr} shows the pair correlation for different values
of $\delta$. For $\delta <5$ mm,  the pair correlation shows a basically 
uncorrelated activity, except at short range. 
For $\delta=5.26$ mm,  long-range pair correlations start to emerge and finally, for 
larger $\delta$, the long-range pair correlations become even stronger. In the limit of very large
$\delta$, stationary states emerge in which the system freezes into a single
fully-ordered state, with all nodes states either $+1$ or $-1$. As mentioned above
in the fully-ordered state with no noise $\alpha$ is infinite.

Our analysis confirms that there is indeed a intermediate scaling regime in the
organization of the brain patterns which reminds fluid turbulence, yet with
a higher degree of spikiness.
Indeed, by sticking to the physiological value $\delta=5.55$ mm, the 
Hopfield brain appears more irregular (lower scaling exponent) than a random walk and 
also of homogeneous incompressible turbulence. 
The most important feature, though, is the sigmoidal dependence  
of the relation $\alpha(\delta)$ with the rampup region 
between about $4$ and $7$ mm, centered around $\delta \sim 5.9$.
This is indeed a physiologically relevant scale of local regions in the
brain. Furthermore, this rampup region unveils the major sensitivity of the scaling exponent to small changes 
of the decay length, indicating that small changes of the decay length lead to fairly
different scaling regimes. In particular, for $\delta$ below $4$ mm, the activity appears to 
be nearly chaotic, whereas, for $\delta$ above $8$ mm, it becomes kind of ``rigid'', \emph{i.e.} 
globally correlated. 

\section{Thresholding the connections}

Having assessed that delocalization (large $\delta$s) promotes a correlated 
response, it is natural to inquire about the role of short versus long-range couplings 
in promoting the correlated patterns sustaining the scaling regime.

To this purpose, we performed a series of simulations by progressively removing
pairs of nodes whose coupling strengths lie below a given 
threshold, namely $J_{ij} < J_{th}$, where the coupling strengths vary in the range $[0,1]$.
Since $J_{ij}$ decays exponentially with the pair distance
a threshold $J_{th}$ excludes pairs beyond a distance 
\begin{equation}\label{eq:dth}
d_{th}=\delta ln{(1/J_{th})},
\end{equation} 
thereby retaining only a fraction 
$1-J_{th}$ of the full set of interacting nodes.   
Thus, to a certain dilution corresponds a certain cutoff distance above
which all couples are disconnected.
The result of this progressive dilution on the scaling exponent are reported in Figure \ref{fig:dilution}. 
Here the dilution $\rho$ is defined as the fraction of disconnected nodes
$i,j$, $J_{ij}=0$, and the result on the scaling exponent is measured comparing the undiluted
exponent $\alpha(0)$ with the exponent $\alpha(\rho)$ at that dilution. 

\begin{figure}[h!]
\centering
\includegraphics[width=.6\columnwidth]{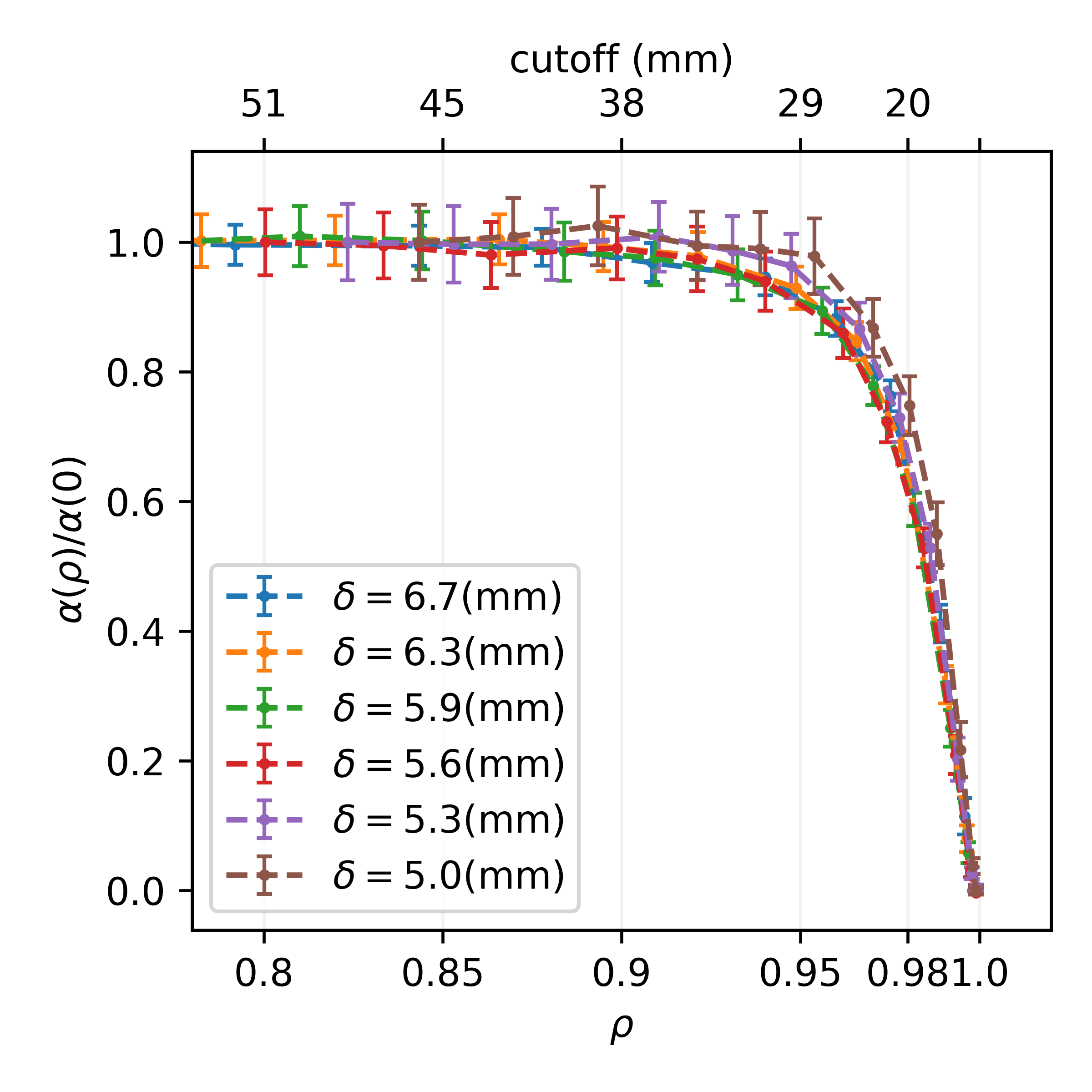}

\caption{\label{fig:dilution} 
The average slope of $\log S(d)$ for different dilution levels $\alpha(\rho)$,
normalized to its threshold-free value $\alpha(0)$ at that given $\delta$.
The dilution is measured as the fraction of disconnected pairs $i,j$, 
$J_{ij}$ elements with value 0.
The second $y$-axis on top shows the cutoff distance in mm for which at a certain
dilution $\rho$ the pairs at a larger or equal distance are disconnected.
The differently colored curves indicate simulations with different $\delta$ values.
The graph shows that within the measurement error, the system is not affected
by the dilution of the edges, up to the removal of more than the $95\%$ of the connections,
in descending order of distance.  
This shows that, in view of the exponential connectivity $J_{ij} = e^{ -d_{ij}/\delta}$,  the
onset of  collective patterns is sustained mostly by the close connections,  within about $4 \delta$.
}
\end{figure}

From this figure, a smooth trend at increasing $\rho$ is observed, with
the scaling exponent starting to display a significant decrease around
$\rho= \left[0.95, 0.98\right]$, when just 2 or 5 percent of the nodes are left, corresponding
to $d_{ij} > 4.6 \delta$. 
With $J_{th}=0.10$, corresponding to $d_{ij} > -\delta ln(0.1) \sim 0.46 \delta$, the 
scaling exponent is basically halved, indicating a major loss of correlation.
This is expected since $J_{th}=0.10$ cuts out all but shortest range interactions.
The conclusion of this dilution analysis is that the brain response is carried almost
entirely by interactions up to about five decay lengths, namely about $1/6$ of the
size of the global brain.
This sounds reasonable, as it strikes a plausible compromise between
short-range ($d< 4\delta$) and long-range ($d > 6 \delta$) interactions.  

It is interesting to observe that a ``solid'' (strongly correlated)  brain, such 
as the one that results for $\delta> 10$ mm, and low dilution $\delta$, would be too 
homogeneous to be able to store and process the amount of information 
required to function properly.
On the opposite side, for too short connectivity length, the brain would 
behave like a ``gas'' of disconnected nodes, hence incapable of collective 
behavior, which is key to its proper functioning.
It appears like the physiological length is achieving an optimal compromise 
between these two opposite, order-disorder, trends.  
With a daring but conducive metaphor, we could speculate
that the brain works in a sort of turbulent liquid-like state,  although 
more spiky than actual turbulent fluids. 

\section{Changing The Number of Parcels}

\begin{figure}[h]
\centering
\includegraphics[width=.7\columnwidth]{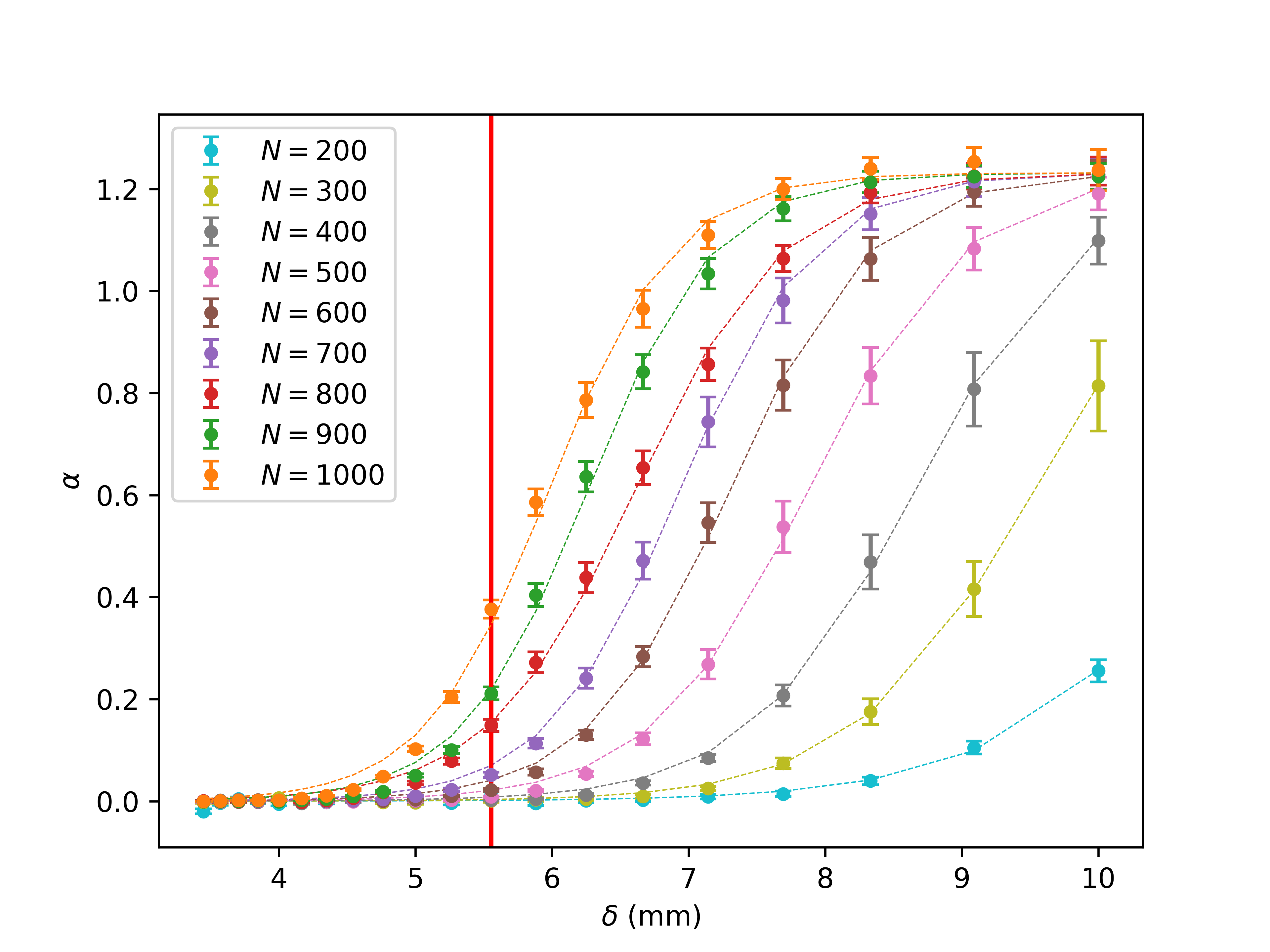}

\caption{\label{fig:dilution} 
                \label{fig:vsN} 
                The average slope of $\alpha$ as a function of $\delta$
                for Schaefer 2018 Parcellations with different sizes $N$ \citep{Schaefer2018}.
                Given each size $N$, we fitted the points with the sigmoid functions in Eq. (\ref{eq:exp}).
}
\end{figure}

To assess the impact of the parcellation size on the scaling picture discussed in this 
paper, we gathered Schaefer 2018 Parcellations \citep{Schaefer2018} with different sizes $N$ and we run
numerical simulations with $N_r = 1000$ replicas and different values of $\delta$.
This allowed us to inspect the relation between $\alpha$ and $\delta$ 
for different values of the size $N$.
The result is the relation reported in Fig. \ref{fig:vsN}, which shows a progressive regression of the
transition regions towards smaller values of $\delta$, combined with a corresponding steepening
of the transition. 

To provide a quantitative measure of this process, we fitted the points at 
given values of $N$ with the sigmoid function:
\begin{equation}\label{eq:exp}
\alpha(\delta) = \frac{\alpha_{\infty}}{1 + \exp{(-k (\delta-\delta_o))} }. 
\end{equation}
where $\alpha_{\infty}$ represents the plateau value of the scaling exponent at 
infinite decay length, $\delta_0$ is the transition scale and $k$ is the steepness of
the transition region. 

The sigmoid function captures the functional relation of the points reasonably well
even though, as we discussed earlier on, the values $\alpha(\delta)$ do not fully converge 
in the top branch because for $\delta > 10$ mm fully aligned configurations prevail, 
with correspondingly divergent values of $\alpha$. Thus, we fitted the function in three steps: first, we estimated $\alpha_{\infty}$ on the plateau points  
for $N = 800, 900, 1000$. Then, we obtained $\delta_0$ and $k$ by fitting the sigmoid 
function $\alpha(\delta)$ on all points using a nonlinear least squares fit. 
We finally obtained $\delta_0$ and $k$ as a function of $N$ via a linear least square fit 
in loglog scale, and found
$\delta_0 \sim N^{-0.379}$ and $k \sim N^{0.328}$, with a $R^2$-value $0.999$ and $0.925$, respectively.

Based on the above scaling of the parameters, one would conclude that the sigmoid 
tends to a sign function centered in zero in the thermodynamic limit $N \to \infty$. 
In plain words, the transition length goes to zero and the transition region becomes 
infinitely steep. Hence, in the infinite $N$ thermodynamic limit, the system is fully correlated (``rigid'') 
even for vanishing decay lengths, an extreme instance of long-range ordering. 

But the brain is not infinite, hence it is plausible to expect that there 
should exist optimal parcellation size $N_{opt}$ providing a correspondingly
optimal Hopfield mass model approximation to the empirical correlations provided by a 
realistic model of structural connectivity.
Due to its inherently coarse-grained nature, above $N_{opt}$, the Hopfield network 
whole brain mass model is no longer expected to provide a trustworthy 
approximation of the collective brain dynamics. 

Even if we cannot pin down such an optimal value, we expect it to lie between $N=10^3$, the 
largest parcellation presently available to us and $N<10^{6}$, which provides a plausible 
upper bound to the number of functioning ``modules'' in the human brain 
(number of neurons, $10^{11}$, divided by the typical module size, $10^{5}$ neurons).
 
Given the actual scalings provided by our fit, a three orders of magnitude increase in the parcellation
size would lead to about ten times smaller values of both $\delta_0$ and thickness $1/k$, thus
placing the transition length $\delta_0$ at about $0.5$ mm, with a 
strong sensitivity of the scaling exponent confined within a region 
of lengthscale about $0.1$ mm wide. 
In other words, going to larger parcellations makes the scaling exponent 
increasingly sensitive to small changes of the decay length.

Future experimental developments will allow us to explore finer parcellations 
with larger $N$ values and test the predictions made in this work. 
Furthermore, from these and other independent 
considerations \cite{kaiser2014hierarchical}, it is plausible to speculate that 
for $N>10^3$,  the connectivity may no longer be approximated 
by a decaying exponential. 

All of the above, point to the importance of obtaining larger 
parcellations, providing further constraints to improve 
future mesoscale models of the global brain activity.

\section{Conclusions}

In conclusion, we have shown that an Hopfield recurrent neural network, informed by experimentally 
derived brain topology, recovers the scaling  picture recently proposed 
by 
\cite{Deco2020}, according to which
the process of information exchange within the human brain shows spatially correlated patterns
qualitatively similar to those displayed by turbulent flows.
Our analysis confirms the initial finding by \cite{Deco2020} in near-quantitative form, 
predicting a very similar scaling exponent ($2/5$ instead of $1/2$) for the same value of the
connectivity decay length. Given the very different 
nature of the two mathematical models, this provides  
a significant hint at the robustness of the scaling picture. 

It is further observed that the scaling exponents are smaller than for the case 
of turbulence, indicating that the collective activity of the brain
is more irregular (spikier) than homogeneous incompressible turbulence.  

The scaling exponents show a steep dependence on the spatial
range of neural connections, hence the specific value of the decay length
has a major impact on the degree of irregularity (spikiness) of the collective patterns.
Such sensitivity is also predicted  to increase by increasing the size
of the parcellation.
This provides a strong indication towards the existence of an optimal connectivity
length $\delta_0$ at which the brain shows non-smooth functional behavior, meaning
that longer/shorter connectivity lengthscales would lead to 
excessive smoothness/roughness respectively. 
With $N=10^3$ parcels, this lengthscale is about 
$\delta_0 \sim 5.5$ mm, well below the global size of the brain. 
For larger parcellations, both the transition length $\delta_0$ and the width
of the transition region are predicted to decrease approximately like
$N^{-1/3}$, which highlights the importance of assessing the optimal parcellation size.

Finally, a dilution analysis shows that, again with $N=10^3$, the Hopfield model brain 
remains functional after removing links above about six decay lengths, 
namely $25-30$ mm, corresponding to about one-fifth of the size of the global brain.
This indicates that, in terms of connectivity decay length, the brain seems to function in a sort
of intermediate ``turbulent liquid-like'' state, whose essential connections lie in near geometrical
mean between the decay length and the global size of the brain.


\section{Acknowlegments}

SS wishes to acknowledge financial support 
from the Bertarelli Foundation during his stay at the Neurobiology Department
of the Harvard Medical School.  
GG wishes to acknowledge financial support H2IOSC Project - Humanities and cultural Heritage Italian Open Science Cloud funded by the European Union – NextGenerationEU – NRRP M4C2 - Project code IR0000029 - CUP B63C22000730005.
The authors wish to acknowledge illuminating discussions with John A. Assad, Haim Sompolinsky, and Bernardo Sabattini.


\bibliographystyle{elsarticle-harv} 
\bibliography{turboBiblio}

\end{document}


\begin{frontmatter}



\title{Supplementary Material: Evidence of Scaling Regimes in the Hopfield Dynamics of Whole Brain Model} 


\author[a,b]{Giorgio Gosti} 
\ead{giorgio.gosti@cnr.it}

\affiliation[a]{organization={Center for Life Nano- and Neuro-Science, Istituto Italiano di Tecnologia},
            addressline={Viale Regina Elena 201}, 
            city={Rome},
            postcode={I-00161}, 
            country={Italy}}
            
\affiliation[b]{organization={Istituto di Scienze del Patrimonio Culturale, Consiglio Nazionale delle Ricerche},
            addressline={Strada della Neve s.n.c., Via Salaria km 29.300}, 
            city={Montelibretti (RM)},
            postcode={I-00010}, 
            country={Italy}}
            
\author[a,c]{Sauro Succi}

\affiliation[c]{organization={Istituto per le Applicazioni del Calcolo del Consiglio Nazionale delle Ricerche},
            addressline={via dei Taurini 19}, 
            city={Roma},
            postcode={I-00185}, 
            country={Italy}}

\author[a,d]{Giancarlo Ruocco\corref{cor1}}
\ead{giancarlo.ruocco@roma1.infn.it}

\affiliation[d]{organization={Dipartimento di Fisica, Universit\'a di Roma ``La Sapienza''},
            addressline={P.le Aldo Moro 5}, 
            city={Roma},
            postcode={I-00185}, 
            country={Italy}}

\cortext[cor1]{Corresponding author}

\end{frontmatter}

\setcounter{equation}{0}
\renewcommand{\theequation}{S.\arabic{equation}}

\section{The turbulent-like whole-brain model}.

Deco \emph{et al.} [1] model the global human brain as a set of $N$ coupled Hopf
oscillators of the form
\begin{eqnarray}
\frac{dx_i}{dt} = a_i x_i &+&(\beta y_i-x_i)(x_i^2+y_i^2)  -\omega_i y_i  \nonumber \\
&+& G \sum_{j=1}^N  J_{ij} (x_j-x_i) - \nu_i \xi_i  \\
\frac{dy_i}{dt} = a_i y_i &+&(\beta x_i+y_i)(x_i^2+y_i^2)  -\omega_i x_i  \nonumber\\
&+& G \sum_{j=1}^N  J_{ij} (y_j-y_i) - \nu_i\xi_i 
\end{eqnarray} 

In the above, $(x_i,y_i)$ are the real and imaginary part of the BOLD signal recorded at the $i$-th node
of the network, $a_i=-0.02$ are decay terms at the brink of the Hopf bifurcation, where scaling is best observed,
$\omega_n$ is the natural frequency of the oscillators, $\beta$ the so called shear factor  and $\xi$
a gaussian noise with unit variance.  
The global coupling across the network is represented through
the connectivity matrix $J_{ij}$ which is assumed to obey the Exponential Distance Rule
\begin{equation}
\label{EDR}
J_{ij} = e ^{-d_{ij}/\delta}
\end{equation}   
where $d_{ij}$ is the euclidean distance between nodes $i$ and $j$
and we have set $\delta$ ($\equiv 1/\lambda$ in Deco's notation).
[1] argues that the exponential distance rule matches very well 
the empirical HCP dMRI tractography of the human brain.
In particular, it reproduces the decay of the fiber densities between 
the parcel centroid pairs as a function of the Euclidian distance. 
The authors [1] further show 
that the optimal fit is obtained for $\delta = 5.55 $ \si{mm}.

\section{Amplitude of turbulence and scaling regime}

In order to assess the scaling properties of the ``turbulent brain'' (TB), 
Deco \emph{et al.} introduce the
following complex Kuramoto Order Parameter:
\begin{equation}
\label{KOP}
\Psi_i(t) \doteq R_i(t) e^{i \theta_i(t)} \doteq \sum_{j=1}^N W_{ij} e^{i \phi_i(t)}
\end{equation} 
where $W_{ij} = J_{ij}/\sum_{k=1}^N  J_{ik}$
and $\phi_i(t)$ is the phase of the $i$-th oscillator obtain by Hilbert-transform 
of the (narrow band) BOLD signal. 
The authors then measure the amplitude of turbulence as the variance of 
the fluctuations of amplitude of the order parameter, namely: 
\begin{equation}
D \doteq <R^2> - <R>^2
\end{equation}
where bracket indicate average in space and time over the entire network 
and over hundreds of noise realizations.

The scaling regime is assessed by inspecting the associated structure function
\begin{equation}
S_2(d)= < (R(x+d)-R(x))^2> \sim d^{\alpha}
\end{equation}
where ``$x$'' and ``$x+d$'' indicate that the two network nodes are at distance $d$ and the average 
is taken at statistical steady state, over all network nodes and noise realizations.
This observable is commonly used to assess the statistical properties of fluid turbulence, and
for the case of Kolmogorov homogeneous incompressible turbulence one obtains
$$\alpha_T = 2/3,$$ indicating that the change in the (longitudinal) flow field
at a distance $d$, $\delta u(d) = \langle |u(x+d)-u(x)|\rangle $, scales like $d^{1/3},$ hence the turbulent
flow field is not differentiable (sometimes $\alpha$ is also called the roughness exponent) [2,3].
For the whole-brain model, Deco \emph{et al.} 
[1] report $$\alpha_D \sim 1/2,$$indicating that the correlation 
pattern of the brain signals is {\it less} regular than those observed in homogeneous 
incompressible turbulence. 

As we shall show, these values should be taken with extreme caution, as they are
very sensitive to a number of assumptions, primarily the structure and functional dependence
of the connectivity matrix on the internode distance. 

\section{Structural connectivity}

We determined the couplings matrix $J_{ij}$ using Schaefer's cerebral 
cortical parcellation atlas. This parcellation was obtained from the analysis of a large dataset 
with a sample size of 1489 [30]. 
This dataset is publicly available, we 
used Schaefer's 1000 parcellation with parcel centroids mapped onto 
the MNI152 volumetric space\footnote{Github repository dataset url [30]: \url{https://github.com/ThomasYeoLab/CBIG/tree/master/stable_projects/brain_parcellation/Schaefer2018_LocalGlobal/Parcellations/MNI/Centroid_coordinates}}.
Fig. 1A shows the spatial distribution of the centroids in the MNI152 volumetric space.
From the centroids of these areas, we computed the pairwise distances. 
Fig. 1B  shows the histogram of the pair distances $d_{ij}$ between each couple of centroids. 
Figures 1C and D show respectively the matrix composed of the Euclidean distances $d_{ij}$, the matrix obtained with the threshold $J_{ij}>0.4$ were $J_{ij} = e^{-d_{ij}/\delta}$ with $\delta$=0.55.

%

\clearpage

\begin{figure}[t] 
\begin{center}
\includegraphics[width=1.\columnwidth]{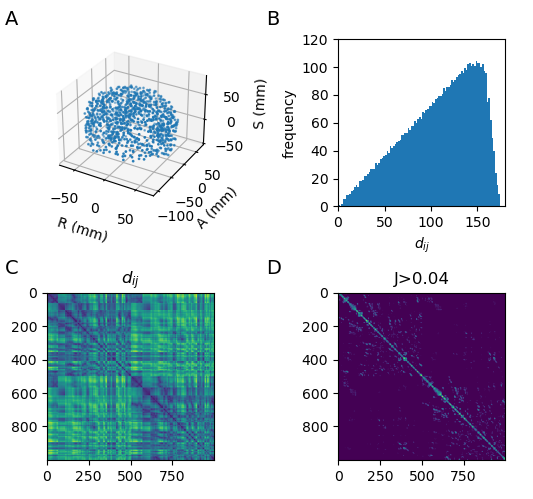}
\end{center}

\noindent {\bf Fig. 1.} 
A) 1000 parcel centroids in three-dimensional space from Schaefer's cerebral cortical parcellation atlas [30].
B) Frequency histogram of the Euclidean distances $d_{ij}$ between 
each pair of nodes from Schaefer's cerebral cortical parcellation atlas [30]. 
The linear dependence on $d_{ij}$ indicates that the nodes are basically distributed 
on a two-dimensional surface.
C) Matrix composed of the Euclidean distances $d_{ij}$ between 
each pair of nodes from Schaefer's cerebral cortical parcellation atlas [30].
D) Matrix formed by the values such that $J_{ij}>0.4$, where $J_{ij} = e^{-d_{ij}/\delta}$ with $\delta$=0.55.

\end{figure}